\documentclass{appolb_mod} 
\usepackage{graphicx}
\usepackage{mathrsfs}
\usepackage{amsmath}
\usepackage{amssymb}
\usepackage{hyperref}

\begin{document}

 \title{How to determine the branch points of correlation functions in Euclidean space %
 \thanks{Excited QCD 2013, 3--9 Feb. 2013, Bjelasnica Mountain, Sarajevo, Bosnia-Herzegovina}%
 }
 \author{Andreas Windisch$^\dagger$, Markus Q. Huber$^\ddag$, Reinhard Alkofer$^\dagger$
 \\
 \address{$^\dagger$ Institut f\"ur Theoretische Physik, Karl-Franzens-Universit\"at Graz, Universit\"atsplatz 5, 8010 Graz, Austria}
 \\
 \address{$^\ddag$Institut f\"ur Kernphysik, Technische Universit\"at Darmstadt, Schlossgartenstrasse~2, 64289 Darmstadt, Germany}
 }
\maketitle
\begin{abstract}
Two-point correlators represented by either perturbative or non-per\-turbative integral equations in Euclidean space are considered. In general, it is difficult to determine the analytic structure of arbitrary correlators analytically. When relying on numerical methods to evaluate the analytic structure, exact predictions of, {\it e.g.}, branch point locations ({\it i.e.}, the multi-particle threshold) provide a useful check. These branch point locations can be derived by Cutkosky's cut rules. However, originally they were derived in Minkowski space for propagators with real masses and are thus not directly applicable in Euclidean space and for propagators of a more general form. Following similar considerations that led Karplus et al., Landau and Cutkosky more than 50 years ago to the mass summation formula that became known as Cutkosky's cut rules, we show how the position of branch points can be derived analytically in Euclidean space from propagators of very general form.
\end{abstract}
\section{Introduction and motivation}
More than 50 years ago, several studies on the analytic structure of Feynman amplitudes were published \cite{Karplus:1958zz,Landau:1959fi,Cutkosky:1960sp} that shed light on the occurrence of the spectral threshold. Here we apply this long-known technique to two-point functions at one-loop order in Euclidean space. Thereby very general analytic structures of the loop integrands are permitted which allows to employ a large class of propagators known in closed form for perturbative calculations, as well as to extend this to non-perturbative equations which need to be solved self-consistently. In principle this is straightforward, but as we applied this technique to various problems in recent studies \cite{Windisch:2012zd,Windisch:2012sz,Windisch:2013mg} and are using it in an ongoing non-perturbative study of the quark propagator, we will provide details of the technique here. The applicability of Cutkosky's cut rules in Euclidean space is discussed also in \cite{Dudal:2010wn} using the inverse Stieltjes transformation. The arguments presented here, on the other hand, are based directly on the loop integral and the analytic structure of the integrand.\par
To start with, consider a two-point function which is expressed in Euclidean momentum space at one-loop order. In the perturbative case, such a two-point function might be written as
\begin{equation}
\mathscr{M}(p)=\langle \mathscr{O}(p)\mathscr{O}(-p)\rangle_{d=4}=\int \frac{d^4q}{(2\pi)^4} \frac{f(p,q,\cos\theta)}{g(p,q,\cos\theta)},
\label{eq1}
\end{equation}
where $d=4$ denotes the (Euclidean) dimension, $q$ is some loop momentum and $\theta$ the angle enclosed by the momenta $p$ and $q$. The non-perturbative case will be addressed in Section \ref{fitted}. For convenience, the integrand is expressed as a fraction of two functions $f$ and $g$, which provides access to the poles of the integrand via the zeros of $g(p,q,\cos\theta)$, assuming that the zeros of $g$ lead to integrable singularities. Note that in general such integrals are divergent and need to be renormalized, {\it e.g.}, via BPHZ. In the following we assume that eq.~(\ref{eq1}) is a renormalized expression.\par
In Euclidean space, the momentum-space integral (\ref{eq1}) might be written in $d$-dimensional hyper-spherical coordinates (here we consider $d=4$, a generalization to arbitrary $d$ is straightforward). Regardless of the dimension $d$, the number of integrals in equation (\ref{eq1}) reduces from $d$ to 2. We thus integrate over the square of the inner momentum $y=q^2$ and $z=\cos\theta$. The expression reduces to
\begin{equation}
\mathscr{M}(x)=\int_0^{\xi^2}dy y\int_{-1}^1 dz\sqrt{1-z^2}\frac{\tilde{f}(x,y,z)}{g(x,y,z)},
\label{eq2}
\end{equation}
with $x=p^2$ and $\xi^2$ is the UV cutoff. Note that the factor of $(2\pi)^4$, as well as the constant contributions of the trivial integrals have been absorbed into the function $\tilde{f}$. When $x\in\mathbb{C}$ instead of $x\in\mathbb{R}^+_0$, a thorough analysis of the analytic structure in the $y$-complex plane allows for the prediction of non-analyticities of the result. The latter we will identify with branch points.\par
When $x$ is not on the positive real axis, the integration in $y$ is no longer straightforward and it becomes necessary to deform the integration contour. As long as it is possible to integrate from $0$ to $\xi^2$ without hitting any obstructions like branch cuts, the result is analytic. Non-analyticities in $x$ arise, when this is prohibited by obstructions in the $y$-plane. These can be poles but also branch cuts induced by the angle integration.\par
\begin{figure}[bt]
\centering
\includegraphics[width=4cm]{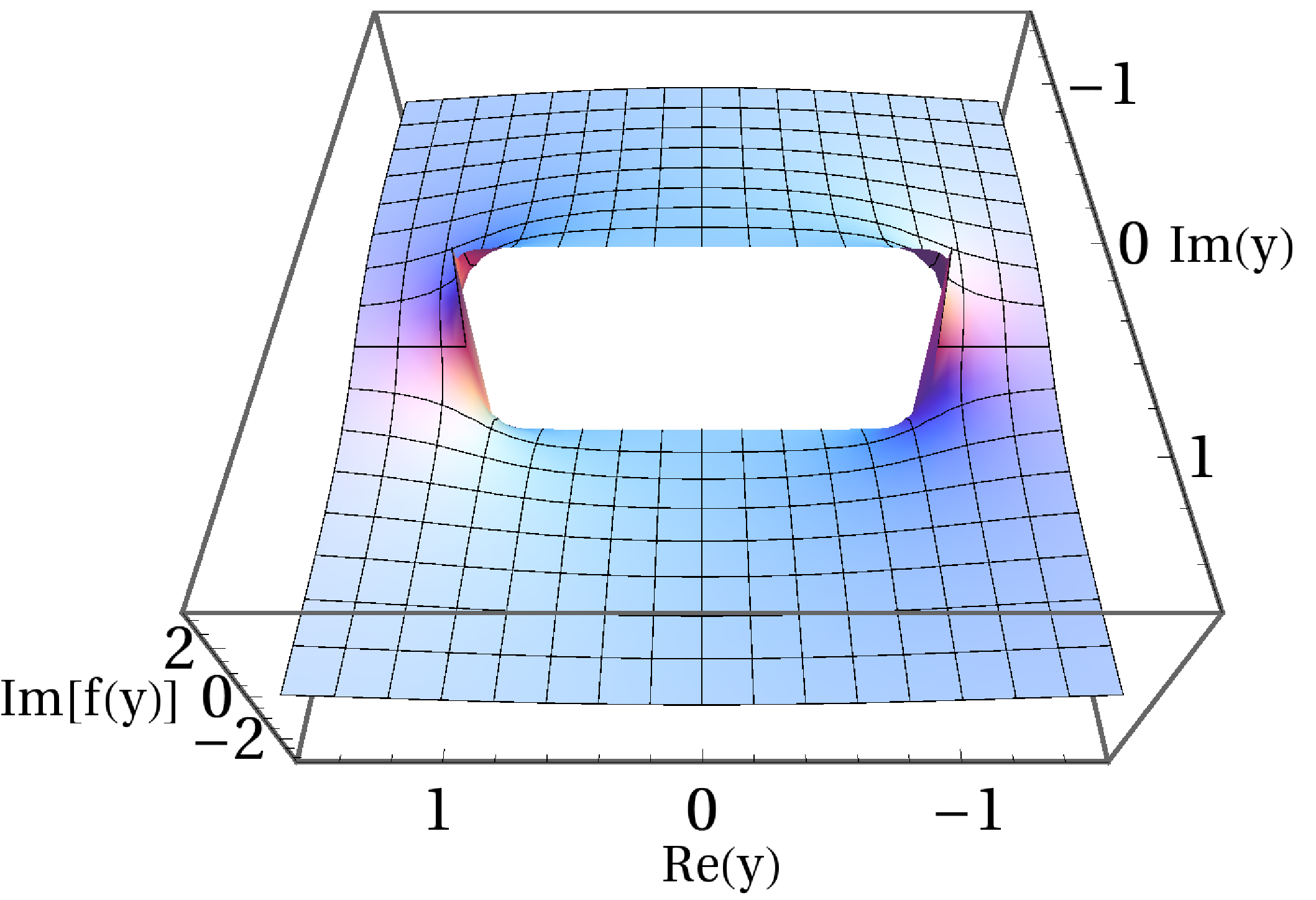}\hspace{2cm}\includegraphics[width=3cm]{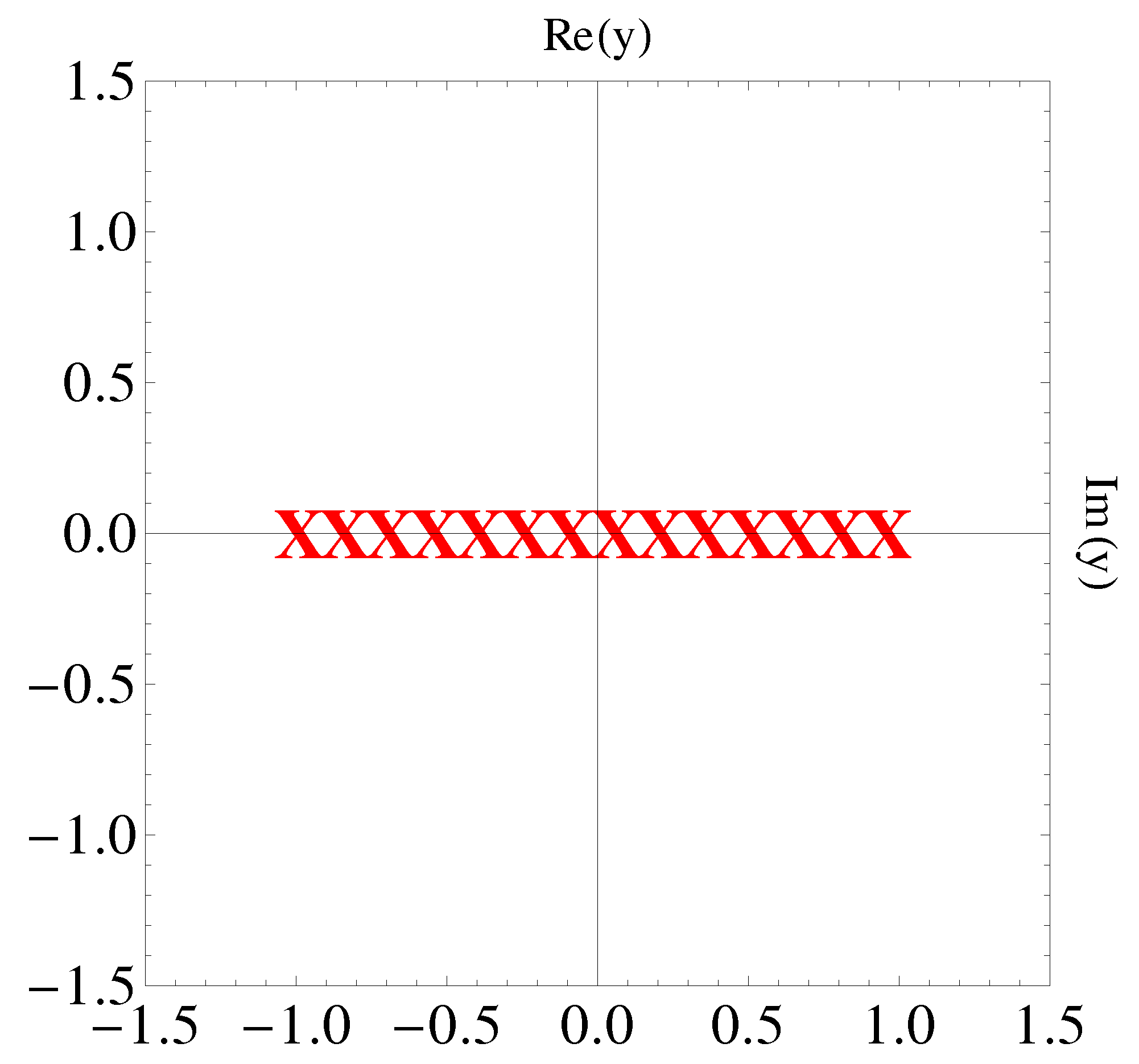}
\label{fig1}
\caption{\label{fig:bc}Branch cut for complex values of $y$ as obtained from equation (\ref{eq3}).}
\end{figure}

The latter arise when integrating over the (integrable) singularities of the angle integral. How this comes about can be illustrated in a very simple example. Consider the following integral for complex values of $y$, where $a,b\in\mathbb{R}$,
\begin{equation}
f(y)=\int_{a}^b dz\frac{1}{y+z}=\log z\bigg|_{y+a}^{y+b}=\log\left(\frac{y+b}{y+a}\right). 
\label{eq3}
\end{equation} 
With the particular (but arbitrary) choice of $a=-1,b=+1$ we find that a branch cut arises for $y\in{[-1,1]}$, see fig.~\ref{fig:bc}. This branch cut is present because $z$ picks up the singular values of the integrand.
This situation is of course much simpler than for physical examples, but it already features a branch cut which is induced by the $z$-integral and shows up in the complex $y$-plane. Furthermore, we learn that we do not have to perform the angular integral in order to predict the branch cut structure in the $y$-plane. It is sufficient to find all complex values of $y$ for which the integrand becomes singular. Thus, in view of equation (\ref{eq2}), the parametrization of the branch cut in the complex $y$-plane for given $x\in\mathbb{C}$ reads  
\begin{equation}
c(x)=y_0(x,z)\bigg|_{z=-1}^{z=+1},
\label{eq4}
\end{equation}
where $y_0(x,z)$ are the zeros of the denominator for given complex $x$, para\-metrized by $z$, {\it i.e.}, $y_0$ is a solution w.r.t. $y$ of 
\begin{equation}
g(x,y,z)=0,\ x\in\mathbb{C},\ x\ \mbox{const.},\ z\in{[-1,1]}.
\label{eq5}
\end{equation}
Solving equation (\ref{eq5}) yields \textit{all} branch cuts in a parametrized form as given in (\ref{eq4}), as well as \textit{all} purely $y$-dependent non-analyticities. In general there are both, poles and cuts. Non-analyticities in $x$ arise when the restrictions imposed on the contour of $y$ do not allow a proper deformation of the contour. These are the points we are after. To identify them we need to find the values of $x$, where such a contour deformation is not possible. This happens when an endpoint of a branch cut coincides with a pole and/or an endpoint of the solely $y$-dependent non-analyticities of the integrand. From this one can also see how the discontinuity at the branch cut develops: Depending on which side of the branch point a point $x$ lies, the integration contour has necessarily a different shape and the value of the integral is not continuous across the branch cut; see fig.~\ref{fig2} for an example.\par
We can summarize the procedure to identify the branch points as follows:
\begin{itemize}
\item \textbf{Step 1:} Find \textit{all} non-analyticities of the $y$-integrand. $\Rightarrow$ Cuts and poles in the $y$-plane.
\item \textbf{Step 2:} Find the points $x_c$ where the endpoints of the induced cuts coincide with other non-analyticities in the $y$-plane which depend only on $y$ $\Rightarrow$ Candidates for branch points.
\item \textbf{Step 3:} Check if the $y$-contour can be deformed properly for any $x_c$ from Step 2. If not, $x_c$ is a branch point $x_b$.
\end{itemize}
\section{\label{real_masses}Correlator with real masses}
\begin{figure}[tb]
\centering
\includegraphics[width=12.5cm]{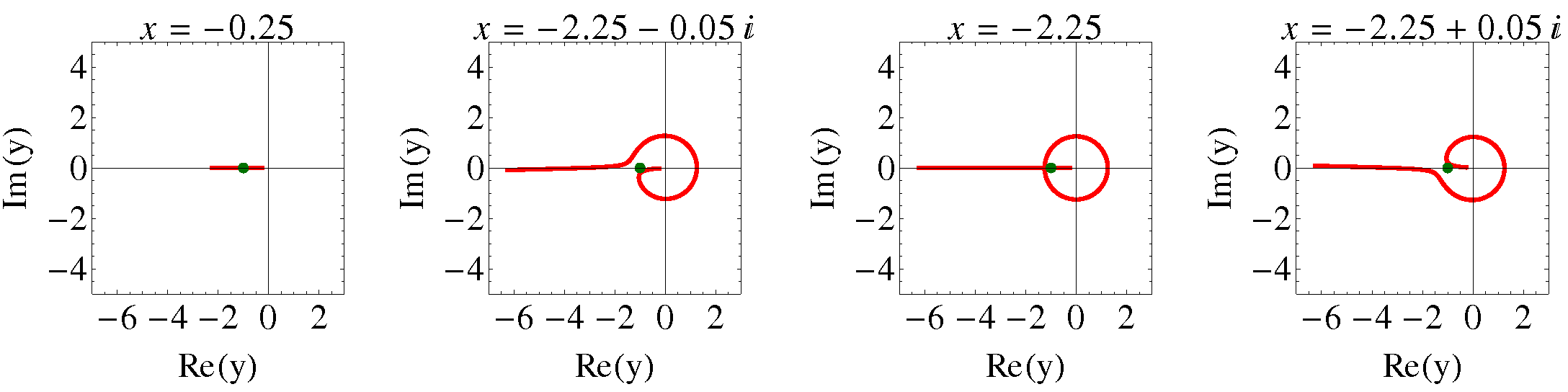}
\caption{The analytic structure in the $y$-plane for $m_1=0.5$ and $m_2=1$. Slightly above/below the predicted branchpoint of $x_0=-(m_1+m_2)^2=-2.25$ the contour can still be drawn, while for $x=x_0$ there is no continuous deformation possible.}
\label{fig2} 
\end{figure}

To confirm the results from the mass-summation formula of \cite{Karplus:1958zz,Landau:1959fi,Cutkosky:1960sp}, let us consider a generic correlator with two masses $m_1\in\mathbb{R^+}$ and $m_2\in\mathbb{R^+}$,
\begin{eqnarray}
\label{generic}
\mathscr{M}(p)&=&\langle \mathscr{O}(p)\mathscr{O}(-p)\rangle_{d=4}=\int\frac{d^4q}{(2\pi)^4}\frac{f(q,p-q)}{((p-q)^2+m_1^2)(q^2+m_2^2)},
\end{eqnarray}
where $f(q,p-q)$ is a function constructed out of the Lorentz invariants. Switching to hyper-spherical coordinates we have
\begin{eqnarray}
\label{corr}
\mathscr{M}(x)&=&\int_0^\infty dy y\int_{-1}^1 dz\sqrt{1-z^2}\frac{\tilde f(x,y,z)}{(x+y-2\sqrt{x}\sqrt{y}z+m_1^2)(y+m_2^2)},
\end{eqnarray}
with the usual identifications $x=p^2$, $y=q^2$ and $z=\cos\theta$. All constants have been absorbed into $\tilde{f}(x,y,z)$. According to the procedure outlined above, we need to determine the zeros of the denominator to get parametrizations for the cuts (Step 1). In this case the solution can be easily worked out analytically and yields
\begin{eqnarray}
\label{sols}
y_1&=&-m_2^2,\\
y_2(x,z)&=&-m_1^2-x+2xz^2-2\sqrt{-m_1^2xz^2-x^2z^2+x^2z^4}\nonumber,\\
y_3(x,z)&=&-m_1^2-x+2xz^2+2\sqrt{-m_1^2xz^2-x^2z^2+x^2z^4}\nonumber.
\end{eqnarray}
According to Step 2 we solve $y_2(x,z)=y_1$ and $y_3(x,z)=y_1$ for $z=\pm 1$ with respect to $x$. This yields the following branch point candidates:
\begin{equation}
x_c=-(m_1\pm m_2)^2.
\label{candidates}
\end{equation}
The solution with the minus sign does not interfere with the integration contour (see fig. \ref{fig2}) so that we are left with
\begin{equation}
x_b=-(m_1+ m_2)^2.
\label{sol}
\end{equation}
This agrees with the result of \cite{Karplus:1958zz,Landau:1959fi,Cutkosky:1960sp} if we use it for Euclidean momenta, {\it i.e.}, if we plug in negative squares of the masses there:
\begin{equation}
x=-(m_1+m_2)^2=\left(i\sqrt{m_1^2}+i\sqrt{m_2^2}\right)^2=\left(\sqrt{-m_1^2}+\sqrt{-m_2^2}\right)^2.
\label{compare}
\end{equation}
For an example with complex masses see \cite{Windisch:2012sz}.
\section{\label{fitted}Fitted correlators and non-perturbative cases: An outlook}
The procedure outlined in the section above proved to be useful in perturbative studies where the analytic continuation was performed numerically. The branch point location(s) obtained from this analysis served as a check for the numerics. In \cite{Windisch:2012zd}, the case of complex conjugate masses was studied, while in \cite{Windisch:2012sz,Windisch:2013mg} complex conjugate masses as well as a propagator featuring a cut with a singular endpoint were considered. For the complex conjugate poles, all three resulting branch points have been found by this procedure. Note that one of them is time-like and thus corresponds to the physical multi-particle threshold, while two are complex. Due to the existence of the latter no K\"all\'en-Lehmann representation is possible.\par
The next step is a self-consistent determination of propagators and their analytic structure in the non-perturbative regime of QCD, {\it e.g.}, as input for bound-state equations. Several studies of the non-perturbative properties of Green functions using Dyson-Schwinger equations (DSEs) have been published, see \cite{Fischer:2009jm,Strauss:2012dg} for recent examples. In an ongoing study we are looking into the analytic properties of the quark propagator DSE in Landau gauge.
\begin{figure}[bt]
\centering
\includegraphics[width=7cm]{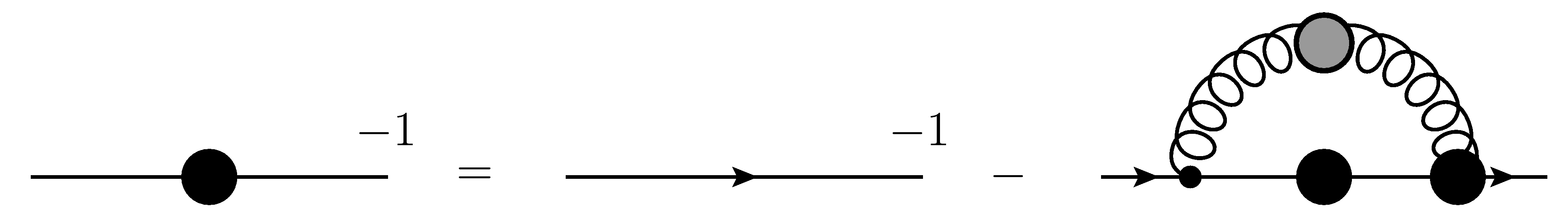}  
\caption{The quark propagator Dyson-Schwinger equation.}
\label{qprop}
\end{figure}
Since the structure of the r.h.s. of the equation depicted in fig. \ref{qprop} is an integral of a similar form as in eq. (\ref{eq2}), the same procedure can be used. The additional complication of such a calculation is that the integral depends on the quantity on the l.h.s. and the equation has to be solved self-consistently.
\section*{Acknowledgments}
We thank Lorenz von Smekal, Christian Fischer, Stefan Strauss, Gernot Eichmann, Richard Williams, Tobias G\"ocke and Walter Heupel for discussions. MQH is supported by the Alexander von Humboldt foundation. AW is funded by the Doctoral Program 'Hadrons in Vacuum, Nuclei and Stars` of the Austrian Science Fund FWF under contract W1203-N16.

\end{document}